\documentclass[11pt,twocolumn,twoside,a4paper,amsmath,amssymb,aps,showkeys,showpacs]{revtex4}

\usepackage{amsfonts}
\usepackage{graphics} %   <------- for LatTeX + EPS
\usepackage{epsfig}
\usepackage{fancyheadings}

\begin{document}
\thispagestyle{myheadings}
%%%%%%%%%%%%%%%%%%%%%%%%%% Title %%%%%%%%%%%%%%%%%%%%%%%%%%%%%%%%%%%%%%
\rhead[]{}%<------
\lhead[]{}%<------
\chead[I.P. Lokhtin, L.V. Malinina, S.V. Petrushanko, A.M. Snigirev, I. Arsene, K. Tywoniuk]{Modeling
the jet quenching in hot and dense QCD matter}%<------short title

\title{Modeling the jet quenching in hot and dense QCD matter}

\author{I.P. Lokhtin, L.V. Malinina, S.V. Petrushanko, A.M. Snigirev}

\affiliation{%
 D.V. Skobeltsyn Institute of Nuclear Physics, M.V. Lomonosov Moscow
	State University, Moscow, Russia }%

\author{I. Arsene}

\affiliation{% 
The Department of Physics, University of Oslo, Norway
  }%
\altaffiliation[Current affiliation: ]{Extreme Matter Institute EMMI, GSI 
Helmholtzzentrum fur Schwerionenforchung GmbH, Darmstadt, Germany}

\author{K. Tywoniuk}
\affiliation{% 
Facultade de Fisica, Universidad de Santiago de Compostela, Santiago de Compostela, Spain
}%

\begin{abstract}
One of the important perturbative (``hard'') probes of hot and dense QCD matter 
is the medium-induced energy loss of energetic partons, so called 
``jet quenching'', which is predicted to be very different in cold nuclear 
matter and in QGP, and leads to a number of phenomena which are already seen in 
the RHIC data on the qualitative level. The inclusion of jet quenching and 
other important collective effects, such as radial and elliptic flows, in the 
existing Monte-Carlo models of relativistic heavy ion collisions is discussed. 
Some issues on the corresponding physical observables at RHIC and LHC energies 
obtained with HYDJET++ model are presented.
\end{abstract}



\maketitle

\renewcommand{\thefootnote}{\fnsymbol{footnote}}
\renewcommand{\thefootnote}{\roman{footnote}}

%*****************   The Body of the Article:   *************************

\section{Introduction}
\label{introduction}

One of the basic tasks of modern high energy physics is the study of the 
fundamental theory of strong interaction (Quantum Chromodynamics, QCD) in 
new, unexplored extreme regimes of super-high densities and temperatures through  
the investigation of the properties of hot and dense multi-parton and multi-hadron systems produced in high-energy nuclear 
collisions~\cite{Hwa:2004yg,d'Enterria:2006su,BraunMunzinger:2007zz}. Indeed, 
QCD is not just a quantum field theory with an extremely rich dynamical content 
(such as asymptotic freedom, chiral symmetry, non-trivial vacuum topology, 
strong CP violation problem, colour superconductivity), but perhaps the only 
sector of the Standard Model, where the basic features (as phase diagram, 
phase transitions, thermalisation of fundamental fields) may be the subject 
to scrutiny in the laboratory. The experimental and phenomenological study of 
multi-particle production in ultrarelativistic heavy ion 
collisions is expected to provide valuable information on 
the dynamical behaviour of QCD matter in the form of a quark-gluon plasma 
(QGP), as predicted by lattice calculations. 

Experimental data, obtained from the Relativistic Heavy Ion Collider (RHIC) at 
maximum beam energy in the center of mass system of colliding ions $\sqrt{s}=200$ GeV 
per nucleon pair, supports the picture of formation of a strongly interacting 
hot QCD matter (``quark-gluon fluid'') in the most central Au+Au (and likely Cu+Cu) 
collisions~\cite{brahms05,phobos05,star05,phenix05}. This appears as    
significant modification of properties of multi-particle production in heavy
ion collisions as compared with the corresponding proton-proton (or peripheral
heavy ion) interactions. In particular, one of the important perturbative 
(``hard'') probes of QGP is the medium-induced energy loss of energetic 
partons, so called ``jet quenching''~\cite{Baier:2000m,Wiedemann:2009sh}, which is predicted to 
be very different in cold nuclear matter and in QGP, and leads to a number of 
phenomena which are already seen in the RHIC data on the qualitative level, such
as suppression of high-$p_T$ hadron production, 
modification of azimuthal back-to-back correlations, azimuthal anisotropy of 
hadron spectra at high p$_T$, etc.~\cite{Wang:2003aw,d'Enterria:2009am}. A number of approaches 
to the description of multiple scattering of hard partons in the dense 
QCD-matter have been developed~\cite{Wang:1994fx,Baier:1996kr,Zakharov:1997uu,Gyulassy:2000fs,Wiedemann:2000tf,Wang:2001ifa,
Wang:2001cs,Vitev:2005yg} 
taking into account the interference pattern for the emission of gluons with a finite 
formation time (QCD analog of the Landau-Pomeranchuk-Migdal effect in QED~\cite{lpm1,lpm2})
Besides the radiative loss (which is supposed to be dominant mechanism of medium-induced partonic energy loss), collisional loss due to elastic scatterings were also quantified in different theoretical models~\cite{bjork82,braaten91,Randrup:2003cw,Markov:2003rk,Mustafa:2003vh,Peigne:2005rk,
Zapp:2005kt,Adil:2006ei,Djordjevic:2006tw,Alam:2006qf,Ayala:2007cq}.

On the other hand, one of 
the most spectacular features of low transverse momentum (``soft'') 
hadroproduction at RHIC are strong collective flow effects: the radial flow 
(ordering of the mean transverse momentum of hadron species with corresponding  
mass) and the elliptic flow (mass-ordered azimuthal anisotropy of particle yields 
with respect to the reaction plane in non-central collisions). The development 
of such a strong flow is well described by the hydrodynamic models and 
requires short time scale and large pressure gradients, attributed to strongly 
interacting systems~\cite{Heinz:2004ar}. 

The heavy ion collision energy in Large Hadron Collider (LHC) at CERN will be 
a factor of $~30$ larger then that in RHIC, thereby allows one to probe new 
frontiers of super-high temperature and (almost) net-baryon free 
QCD~\cite{Abreu:2007kv,hp1,hp2,hp3}. The emphasis of the LHC heavy ion data 
analysis (at $\sqrt{s}=5.5$ TeV per nucleon pair for lead beams) will be on the 
perturbative, or hard probes of the QGP (quarkonia, jets, photons, high-p$_T$ 
hadrons) as well as on the global event properties, or soft probes (collective 
radial and elliptic flow effects, hadron multiplicity, transverse energy 
densities and femtoscopic momentum correlations). It is expected 
that at LHC energies the role of hard and semi-hard particle production will be 
significant even for the bulk properties of created matter. 

\section{Jet quenching and flow in Monte-Carlo models of relativistic heavy ion 
collisions}
\label{Models}

Ongoing and future experimental studies of relativistic heavy ion collisions 
in a wide range of beam energies require the development of new Monte-Carlo 
(MC) event generators and improvement of existing ones. Especially for 
experiments which will be conducted at LHC, due to very high parton 
and hadron multiplicities, one needs fast (but realistic) MC tools for heavy 
ion event simulation~\cite{alice1,alice2,cms}. The main advantage of MC technique 
for the simulation of high-multiplicity hadroproduction is that it allows a visual  
comparison of theory and data, including if necessary the detailed detector 
acceptances, responses and resolutions. A realistic MC event generator should  
include a maximum possible number of observable physical effects which are 
important to determine the event topology: from the bulk properties of 
soft hadroproduction (domain of low transverse momenta $p_T < 1$GeV$/c$)  
such as collective flows, to hard multi-parton production in hot and dense 
QCD-matter, which reveals itself in the spectra of high-$p_T$ particles and 
hadronic jets. 

In most of the available MC heavy ion event generators, the simultaneous 
treatment of jet quenching and collective flow effects is absent. For example, 
the popular MC model HIJING~\cite{hijing} includes jet production and jet 
quenching, but it does not include flow effects. The partonic 
energy loss MC models PYQUEN\cite{Lokhtin:2005px}, JEWEL~\cite{Zapp:2008gi}, 
Q-PYTHIA~\cite{Armesto:2008qh}, YaJEM~\cite{Renk:2009hi} and 
MARTINI~\cite{Schenke:2009gb} combine a perturbative final state parton 
shower with QCD-medium effects and simulate the jet quenching in various
approaches, but without considering the soft component of final hadronic 
state. The event generators 
FRITIOF~\cite{fritiof} and LUCIAE~\cite{luciae} include jet production 
(without jet quenching), while some collective nuclear effects (such as string 
interactions and hadron rescatterings) are taken into account in LUCIAE. 
Another MC model THERMINATOR~\cite{therminator} includes detailed statistical 
description of ``thermal'' soft particle production and can reproduce the main 
bulk features of hadron spectra at RHIC (in particular, describe simultaneously the 
momentum-space measurements and the freeze-out coordinate-space data), but it 
does not consider hard parton production processes. There is a number of 
microscopic transport hadron models (UrQMD~\cite{urqmd}, QGSM~\cite{qgsm}, 
AMPT~\cite{ampt}, etc.), which attempt to analyze the soft particle 
production in a wide energy range, however they also do not include in-medium 
production of high-$p_T$ multi-parton states. Another  
heavy ion event generator, ZPC~\cite{zpc}, has been created to simulate parton 
cascade evolution in ultrarelativistic heavy ion collisions. From a physical 
point of view, such an approach seems reasonable for very high beam energies 
(RHIC, LHC). However the full treatment of parton cascades may require
significant amount of CPU run time (especially for LHC).

\section{HYDJET++ model and its applications for RHIC and LHC}
\label{HYDJET}

HYDJET++ event generator~\cite{Lokhtin:2008xi} includes detailed 
treatment of soft hadroproduction as well as hard multi-parton production, and 
takes into account medium-induced parton rescattering and energy loss. The 
heavy ion event in HYDJET++ is the superposition of two independent components: 
the soft, hydro-type state and the hard state resulting from multi-parton 
fragmentation. Note that a conceptually similar approximation has been 
developed in~\cite{Hirano:2004rs,Armesto:2009zi}. HYDJET++ model is the development and 
continuation of HYDJET event generator~\cite{Lokhtin:2005px}, and it  
contains the important additional features for the soft component: resonance decays and 
more detailed treatment of thermal and chemical freeze-out 
hypersurfaces~\cite{Amelin:2006qe,Amelin:2007ic}. The main program HYDJET++ is 
written in the object-oriented C++ language under the ROOT environment~\cite{root}.
The hard part of HYDJET++ is identical to the hard part of Fortran-written HYDJET and it 
is included in the generator structure as a separate directory.When the generation of soft and hard components in each event at given impact parameter $b$ is completed, the event record (information about coordinates and momenta of primordial particles, decay products of unstable particles and stable particles) is formed as the junction of these two independent event outputs. 

The details on physics model and simulation procedure of HYDJET++ can be found in the 
corresponding manual~\cite{Lokhtin:2008xi}. The main features of HYDJET++ model are 
listed only very briefly below.

\subsection{Hard multi-jet production} 

The model for the hard multi-parton part of HYDJET++ event is the same as that 
for HYDJET event generator, and it based on PYQUEN partonic energy loss 
model~\cite{Lokhtin:2005px}. The approach to the 
description of multiple scattering 
of hard partons in the dense QCD-matter (such as quark-gluon plasma) is based on the 
accumulative energy loss via  the gluon radiation being associated with each parton 
scattering in the expanding quark-gluon fluid and includes the interference effect 
(for the emission of gluons with a finite formation time) using the modified radiation 
spectrum $dE/dl$ as a function of decreasing temperature $T$. The model takes into
account radiative and collisional energy loss of hard partons in longitudinally
expanding quark-gluon fluid, as well as realistic nuclear geometry.  

The Fortran routine for single hard nucleon-nucleon 
sub-collision PYQUEN was constructed as a modification of the jet event 
obtained with the generator of hadron-hadron interactions PYTHIA$\_$6.4~\cite{pythia}. 
The event-by-event simulation procedure in PYQUEN includes {\it 1)} generation of 
initial parton spectra with PYTHIA and production vertexes at given impact parameter; 
{\it 2)} rescattering-by-rescattering simulation of the parton path in a dense zone 
and its radiative and collisional energy loss; {\it 3)} final hadronization according 
to the Lund string model for hard partons and in-medium emitted gluons. Then the 
PYQUEN multi-jets generated according to the binomial distribution are included in the 
hard part of the event. The mean number of jets produced in an AA event is the 
product of the number of binary NN sub-collisions at a given impact parameter and the 
integral cross section of the hard process in $NN$ collisions with the minimum  
transverse momentum transfer $p_T^{\rm min}$. In order to take into account the 
effect of nuclear shadowing on parton distribution functions, the impact parameter 
dependent parameterization obtained in the framework of Glauber-Gribov 
theory~\cite{Tywoniuk:2007xy} is used. 

\subsection{Soft ``thermal'' hadron production}

The soft part of HYDJET++ event is the ``thermal'' hadronic state generated on the 
chemical and thermal freeze-out hypersurfaces obtained from the parameterization 
of relativistic hydrodynamics with preset freeze-out conditions (the adapted C++ code 
FAST MC~\cite{Amelin:2006qe,Amelin:2007ic}). Hadron multiplicities are calculated 
using the effective thermal volume approximation and Poisson multiplicity distribution 
around its mean value, which is supposed to be proportional to the number of 
participating nucleons at a given impact parameter of AA collision. The fast soft 
hadron simulation procedure includes {\it 1)} generation of the 4-momentum of a hadron 
in the rest frame of a liquid element in accordance with the equilibrium distribution 
function; {\it 2)} generation of the spatial position of a liquid element and its 
local 4-velocity in accordance with phase space and the character of motion of the 
fluid; {\it 3)} the standard von Neumann rejection/acceptance procedure to account 
for the difference between the true and generated probabilities; {\it 4)} boost of 
the hadron 4-momentum in the center mass frame of the event; {\it 5)} the two- 
and three-body decays of resonances with branching ratios taken from the SHARE 
particle decay table~\cite{share}. The high generation speed in HYDJET++ is achieved 
due to almost 100\% generation efficiency of the ``soft'' part because of the  
nearly uniform residual invariant weights which appear in the freeze-out 
momentum and coordinate simulation. 

\subsection{Validation of HYDJET++ with experimental RHIC data}  

It was demonstrated in~\cite{Amelin:2006qe,Amelin:2007ic} that FAST MC model can 
describe well the bulk properties of hadronic state created in Au+Au collisions 
at RHIC at $\sqrt{s}=200 A$ GeV (such as particle number ratios, low-p$_T$ spectra, 
elliptic flow coefficients $v_2(p_T, b)$, femtoscopic correlations in 
central collisions), while HYDJET model is 
capable of reproducing the main features of jet quenching pattern at RHIC 
(high-$p_T$ hadron spectra and the suppression of azimuthal back-to-back 
correlations)~\cite{Lokhtin:2005px}. Since soft and hard hadronic states in
HYDJET++ are simulated independently, a good description of hadroproduction at 
RHIC in a wide kinematic range can be achieved, moreover a number of 
improvements in FAST MC and HYDJET have been done as compared to earlier 
versions. A number of input parameters of the model can be fixed from fitting 
the RHIC data to various physical observables~\cite{Lokhtin:2008xi}. 

\begin{figure*}
\includegraphics[width=8.4cm]{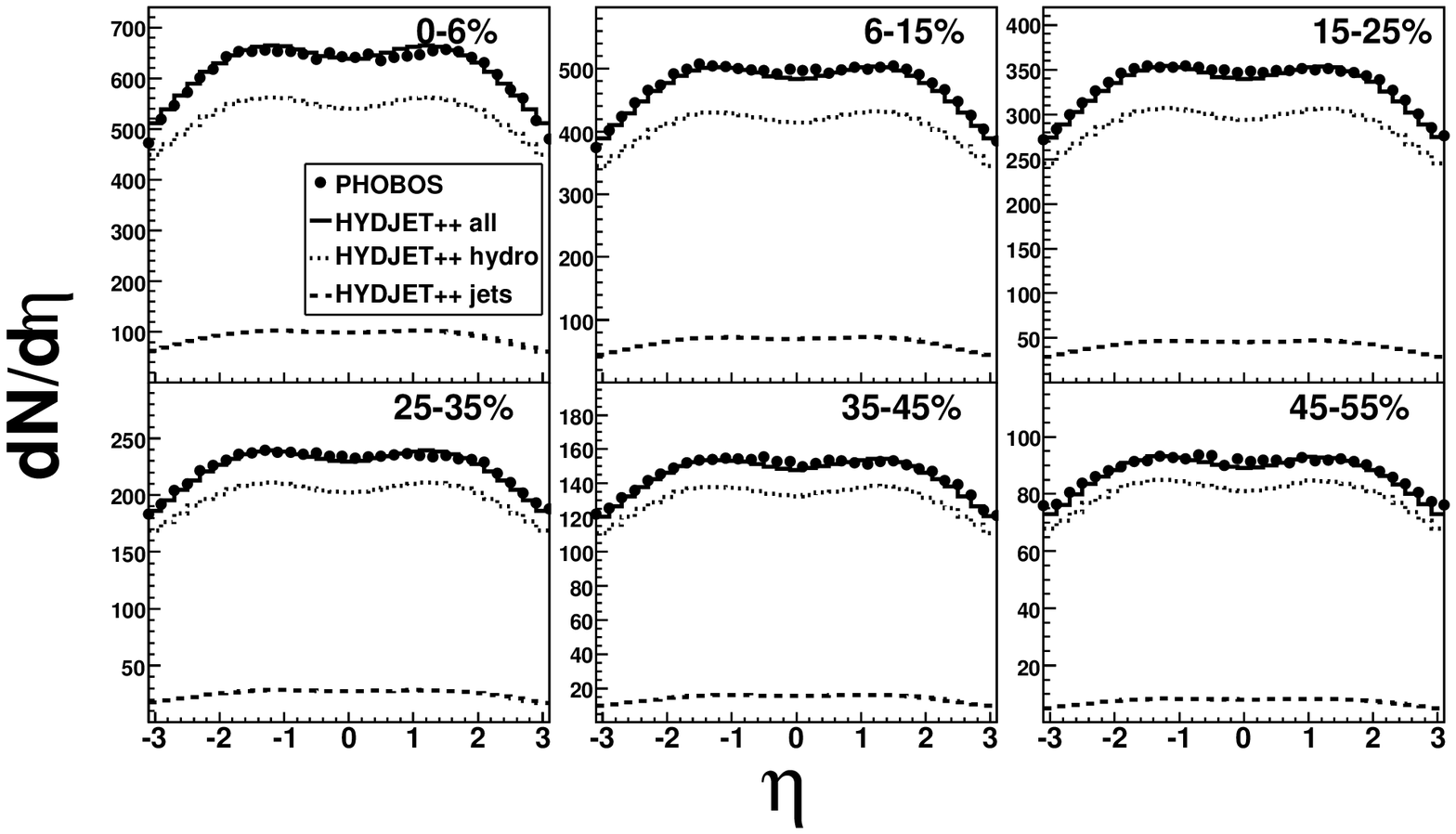} $~$
\includegraphics[width=8.4cm]{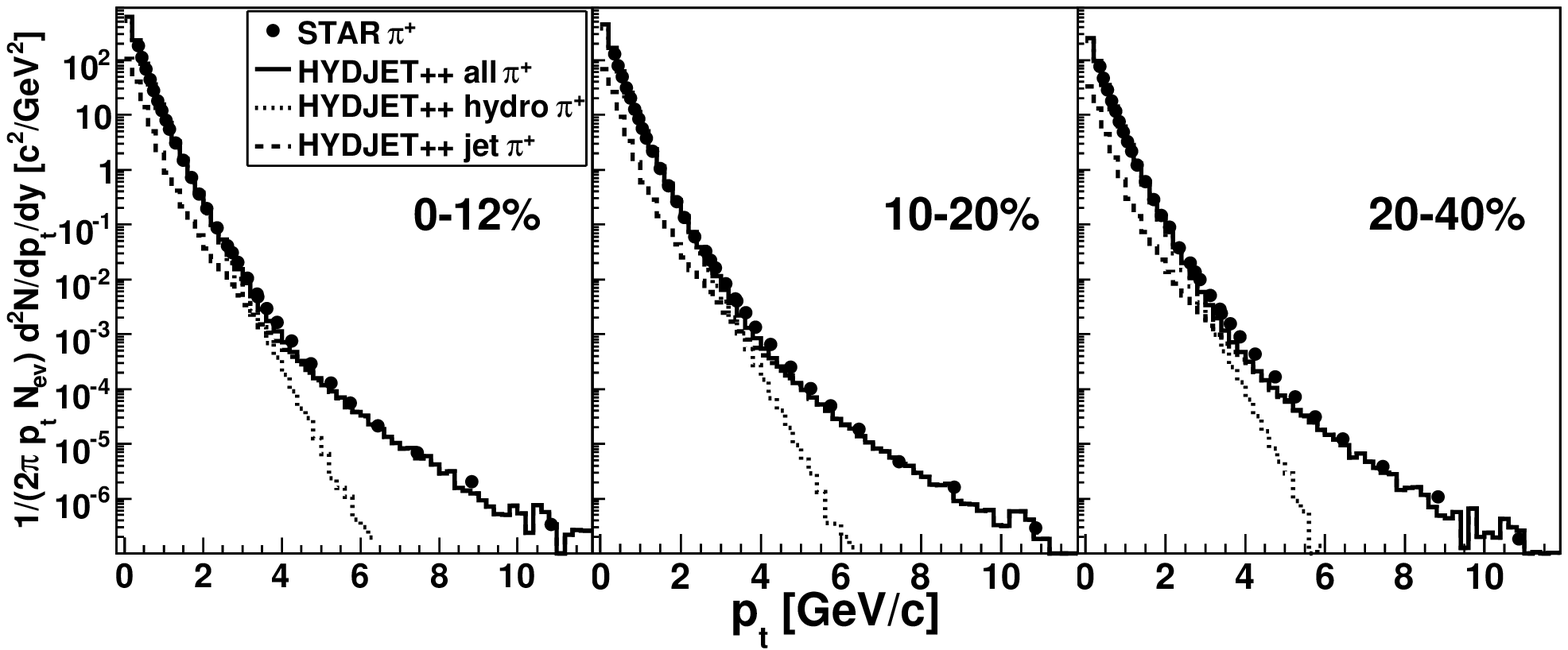}
\caption{The pseudorapidity distribution of charged hadrons (left) and the transverse 
momentum distribution of pions (right) in Au+Au collisions at $\sqrt{s}=200 A$ GeV for 
different centrality sets. The points are RHIC data, histograms are the HYDJET++ calculations 
(solid -- total, dotted -- hydro part, dashed -- jet part).}
\label{hydjet-rhic}
\end{figure*}

\begin{enumerate}  

\item {\bf Ratio of hadron abundances.} It is well known that the particle
abundances in heavy ion collisions in a wide energy range can be reasonable well
described within statistical models based on the assumption that the produced
hadronic matter reaches thermal and chemical equilibrium. The thermodynamical
potentials $\widetilde{\mu_{B}}=0.0285$ GeV, $\widetilde{\mu_{S}}=0.007$ GeV, 
$\widetilde{\mu_{Q}}=-0.001$, the strangeness suppression factor $\gamma_s=1$, 
and the chemical freeze-out temperature  $T^{\rm ch}=0.165$ GeV have been fixed 
in~\cite{Amelin:2006qe} from fitting the RHIC data to  
various particle ratios near mid-rapidity in central Au+Au collisions at 
$\sqrt{s}=200 A$ GeV ($\pi^-/\pi^+$, $\bar{p}/\pi^-$, $K^-/K^+$, $K^-/\pi^-$, 
$\bar{p}/p$, $\bar{\Lambda}/\Lambda$, $\bar{\Lambda}/\Lambda$, $\bar{\Xi}/\Xi$, 
$\phi/K^-$, $\Lambda/p$, $\Xi^-/\pi^-$). 

\item {\bf Low-$p_T$ hadron spectra.} Transverse momentum $p_T$ and transverse 
mass $m_T$ hadron spectra ($\pi^+$, $K^+$ and $p$ with $m_T<0.7$ GeV/$c^2$) near  
mid-rapidity at different centralities of Au+Au collisions at $\sqrt{s}=200 A$ 
GeV were analyzed in~\cite{Amelin:2007ic}. The slopes of these
spectra allow the thermal freeze-out temperature  $T^{\rm th}=0.1$ GeV and 
the maximal transverse flow rapidity in central collisions 
$\rho_u^{\rm max}(b=0)=1.1$ to be fixed.  

\item {\bf Femtoscopic correlations.} Because of the effects of quantum 
statistics and final state interactions, the momentum (HBT) correlation functions 
of two or more particles at small relative momenta in 
their c.m.s. are sensitive to the space-time characteristics of 
the production process on the level of $fm$. The space-time parameters of thermal 
freeze-out region in central Au+Au collisions at $\sqrt{s}=200 A$ GeV have been 
fixed in~\cite{Amelin:2007ic} by means of fitting the three-dimensional correlation 
functions measured for $\pi^+\pi^+$ pairs and extracting the correlation radii 
$R_{\rm side}$, $R_{\rm out}$ and $R_{\rm long}$: $\tau_f(b=0)=8$ fm/$c$, 
$\Delta \tau_f(b=0)=2$ fm/$c$, $R_f(b=0)=10$ fm. 

\item {\bf Pseudorapidity hadron spectra.} The PHOBOS data on $\eta$-spectra of 
charged hadrons~\cite{Back:2002wb} at different centralities of Au+Au collisions 
at $\sqrt{s}=200 A$ GeV have been analyzed to fix the particle densities in the 
mid-rapidity region and the maximum longitudinal flow rapidity 
$\eta_{\rm max}=3.3$ (Fig.~\ref{hydjet-rhic}, left). Since mean ``soft'' and ``hard''
hadron multiplicities depend on the centrality in different ways (they are 
roughly proportional to $\overline{N_{\rm part}(b)}$ and 
$\overline{N_{\rm bin}(b)}$ respectively), the relative contribution of soft 
and hard parts to the total event multiplicity can be fixed through the 
centrality dependence of $dN/d\eta$. The corresponding contributions from 
hydro- and jet-parts are determined by the input parameters 
$\mu_{\pi}^{\rm eff~th}=0.06$ GeV and $p_T^{\rm min}=3.4$ GeV/$c$ respectively.

\item {\bf High-$p_T$ hadron spectra.} High transverse momentum hadron 
spectra ($p_T > 2-4$ GeV/$c$) are sensitive to parton production and 
jet quenching effects. Thus fitting the measured high-$p_T$ tail allows 
the extraction of PYQUEN energy loss model parameters. We assume 
the QGP formation time $\tau_0=0.4$ fm/$c$ and the number of active quark 
flavours $N_f=2$. Then the reasonable 
fit of STAR data on high-$p_T$ spectra of charged pions at different 
centralities of Au+Au collisions at $\sqrt{s}=200 A$ GeV~\cite{Abelev:2006jr} is
obtained with the initial QGP temperature $T_0=0.3$ GeV (Fig.~\ref{hydjet-rhic}, 
right).

\item {\bf Elliptic flow.} The elliptic flow coefficient $v_2$ (which is
determined as the second-order Fourier coefficient in the hadron distribution
over the azimuthal angle $\varphi$ relative to the reaction plane angle
$\psi_R$, so that $v_2 \equiv \left< \cos{2(\varphi-\psi_R)} \right>$) is an
important signature of the physics dynamics at early stages of non-central 
heavy ion collisions. According to the typical hydrodynamic scenario, the
values $v_2(p_T)$ at low-$p_T$ ($< 2$ GeV/$c$) are determined mainly by the 
internal pressure gradients of an expanding fireball during the initial high 
density phase of the reaction (and it is sensitive to the momentum and azimuthal 
anisotropy parameters $\delta$ and $\epsilon$ in the frameworks of HYDJET++), while 
elliptic flow at high-$p_T$ is generated in HYDJET++ (as well as in other jet
quenching models) due to the partonic energy loss in an azimuthally asymmetric 
volume of QGP. The values of $\delta$ and $\epsilon$ were estimated 
in~\cite{Lokhtin:2008xi} for different centrality sets by fitting the measured by 
the STAR Collaboration transverse momentum dependence of the elliptic flow coefficient 
$v_2$ of charged hadrons in Au+Au collisions at $\sqrt{s}=200 A$ 
GeV~\cite{Adams:2004bi}. Note that the choice 
of these parameters does not affect any azimuthally integrated physics 
observables (such as hadron multiplicities, $\eta$- and $p_T$-spectra, etc.), 
but only their differential azimuthal dependences. 

\end{enumerate}

\subsection{Simulations with HYDJET++ at LHC}

The heavy ion collision energy at LHC will be a factor of $30$ larger then that at  
RHIC, thereby allows one to probe new 
frontiers of super-high temperature and (almost) net-baryon free 
QCD. The emphasis of the LHC heavy ion data 
analysis (at $\sqrt{s}=5.5$ TeV per nucleon pair for lead beams) will be on the 
hard probes of the QGP (quarkonia, jets, photons, high-p$_T$ 
hadrons) as well as on the global event properties, or soft probes (collective 
radial and elliptic flow effects, hadron multiplicity, transverse energy 
densities and femtoscopic momentum  correlations). It is expected 
that at LHC energies the role of hard and semi-hard particle production will be 
significant even for the bulk properties of created matter. HYDJET++ seems to be
an effective simulation tool to analyze the influence of in-medium jet 
fragmentation on various physical observables. 

\begin{figure*}
\includegraphics[width=8.4cm]{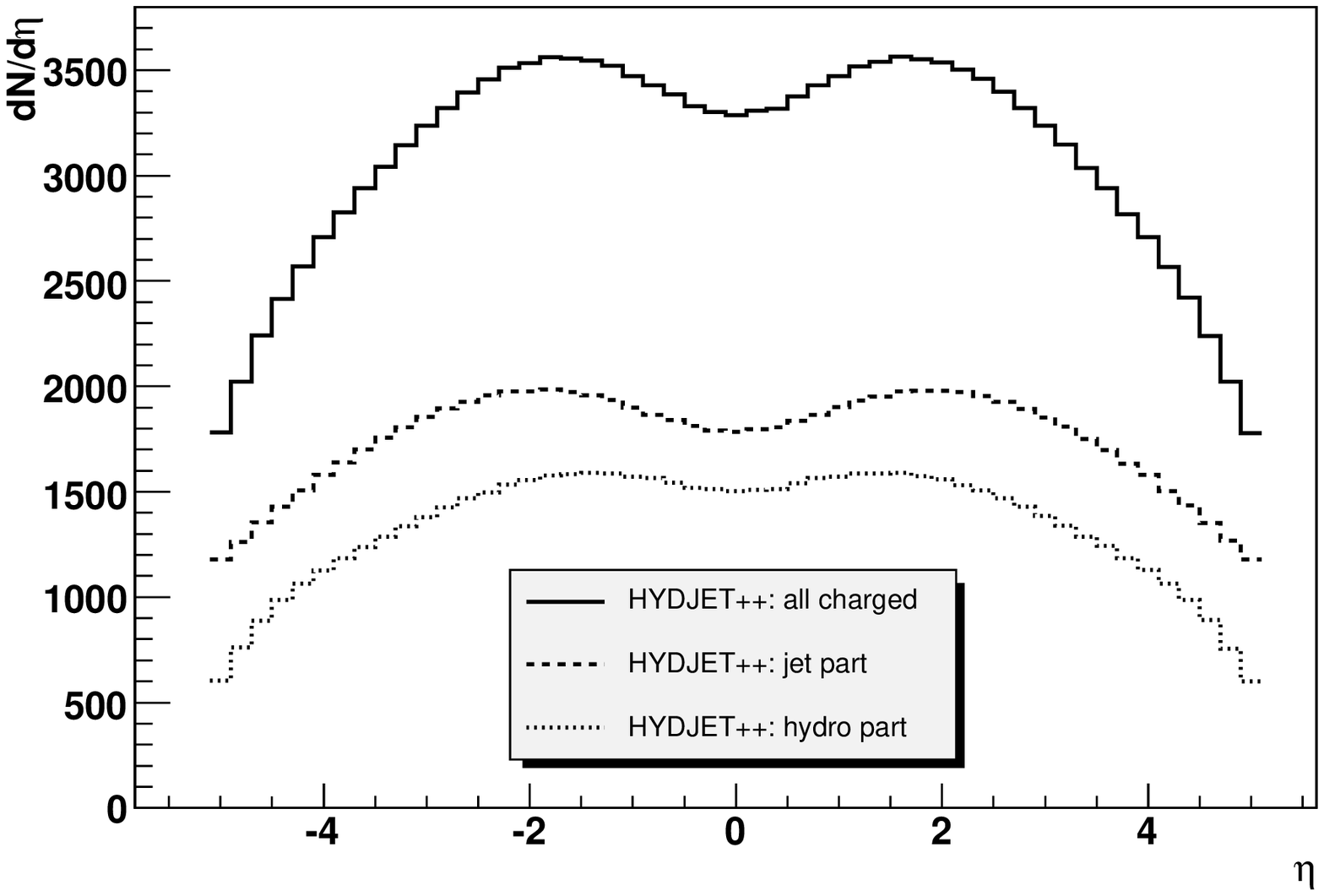} $~$
\includegraphics[width=8.4cm]{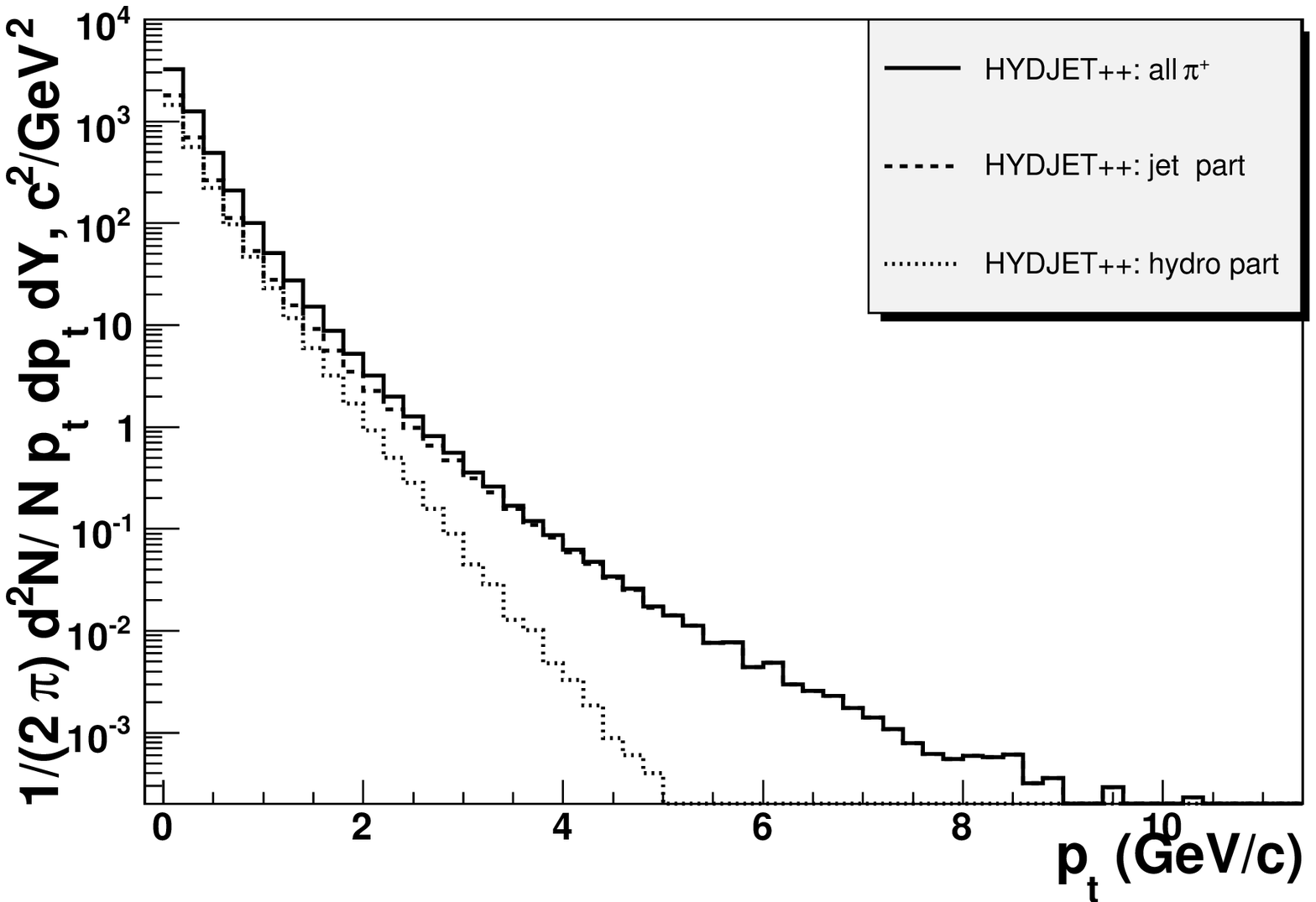}
\caption{The pseudorapidity distribution of charged hadrons (left) and the transverse 
momentum distribution of pions (right) in $5$\% most central Pb+Pb collisions at 
$\sqrt{s}=5500 A$ GeV (solid -- total, dotted -- hydro part, dashed -- jet part), 
$p_T^{\rm min}=7$ GeV/$c$.}
\label{hydjet-lhc}
\end{figure*}

Figure \ref{hydjet-lhc} shows the pseudorapidity distribution of charged hadrons and 
transverse momentum distribution of pions obtained with HYDJET++ 
default settings (in particular, $p_T^{\rm min}=7$ GeV/$c$) for $5$\% most
central Pb+Pb events. The estimated contribution of hard component to the total event 
multiplicity is on the level $\sim 55$\% here, what is much larger than as 
compared with RHIC ($\sim 15$\%). Of course, this
number is very sensitive to the parameter $p_T^{\rm min}$ --- minimal 
$p_T$ of ``non-thermalized'' parton-parton hard scatterings. For example,
increasing the value $p_T^{\rm min}$ up to $10$ GeV/$c$ results in decreasing
this contribution down to $\sim 25$\%. 

Some applications of HYDJET++ for high-p$_T$ studies at the LHC were discussed  
during this Symposium~\cite{malinina-ismd09,bravina-ismd09}. 
In particular, the spectacular predictions of HYDJET++ are possible reducing the 
femtoscopic correlation radii~\cite{Lokhtin:2009be,malinina-ismd09} and elliptic flow~\cite{Eyyubova:2009hh,bravina-ismd09} 
in heavy ion collisions as one moves from RHIC to LHC energies due to the significant contribution of (semi)hard component to the space-time structure of the hadron emission source. 

\section{Summary}
\label{Summary}

One of the important perturbative (``hard'') probes of hot and dense QCD matter 
is the medium-induced energy loss of energetic partons, so called ``jet 
quenching'', which is predicted to be very different in cold nuclear matter and 
in quark-gluon medium, and leads to a number of phenomena which are already 
seen in the RHIC data on the qualitative level. Another spectacular feature of 
hadroproduction at RHIC supporting the pattern of hot QCD matter formation is 
strong collective flow effect (radial and elliptic flows). In order to quantify 
these collective effects in a wide energy range (up to the LHC), one needs the 
realistic Monte-Carlo models. Such event generators should take into account a 
maximum possible number of observable physical effects which are important to 
determine the event topology: from bulk properties of soft hadroproduction 
(domain of low transverse momenta $p_T$) to hard multi-parton production in hot 
QCD-matter, which reveals itself in spectra of high-$p_T$ particles and 
hadronic jets.  

Among other heavy ion event generators, HYDJET++ model focuses on the detailed 
simulation of jet quenching effect basing on the partonic energy loss model 
PYQUEN, and also reproducing the main features of nuclear collective dynamics 
by fast (but realistic) way. The final hadronic state in HYDJET++ represents 
the superposition of two independent components: hard multi-parton 
fragmentation and soft hydro-type part. The soft part of HYDJET++ is the 
``thermal'' hadronic state generated on the chemical and thermal freeze-out 
hypersurfaces obtained from the parameterization of relativistic hydrodynamics 
with preset freeze-out conditions. HYDJET++ is capable of reproducing the bulk 
properties of heavy ion collisions at RHIC (hadron spectra and ratios, radial 
and elliptic flow, femtoscopic momentum correlations), as well as high-p$_T$ 
hadron spectra. HYDJET++ is an effective simulation tool to analyze the 
influence of in-medium jet fragmentation on various physical observables at 
the LHC. 

\section{Acknowledgments} 
\label{Acknowledgments} 
We would like to thank L.V.~Bravina, D.~d'Enterria, G.Kh.~Eyyubova, A.M.~Gribushin, V.L.~Korotkikh, R.~Lednicky, L.I.~Sarycheva, C.Yu.~Teplov, Yu.M.~Sinyukov and E.E.~Zabrodin for joint work and fruitfull discussions. I.L. and L.M. wish to express the gratitude to the organizers of the XXXIX International Symposium on Multiparticle Dynamics for the warm welcome and the hospitality. This work was supported by Russian Foundation for Basic Research 
(grants No 08-02-91001 and No 08-02-92496), Grant of President of Russian Federation 
(No 107.2008.2), Russian Ministry for Education and Science (contracts 01.164.1-2.NB05 and
02.740.11.0244) and Dynasty Foundation.

\label{last}
\end{document}